\documentstyle[apjpt]{article}

\begin{document}

\def\MSUN{\rm M_{\odot}}
\def\RSUN{\rm R_{\odot}}
\def\MSUNYR{\rm M_{\odot}\,yr^{-1}}
\def\MDOT{\dot{M}}

\newbox\grsign \setbox\grsign=\hbox{$>$} \newdimen\grdimen \grdimen=\ht\grsign
\newbox\simlessbox \newbox\simgreatbox
\setbox\simgreatbox=\hbox{\raise.5ex\hbox{$>$}\llap
     {\lower.5ex\hbox{$\sim$}}}\ht1=\grdimen\dp1=0pt
\setbox\simlessbox=\hbox{\raise.5ex\hbox{$<$}\llap
     {\lower.5ex\hbox{$\sim$}}}\ht2=\grdimen\dp2=0pt
\def\simgreat{\mathrel{\copy\simgreatbox}}
\def\simless{\mathrel{\copy\simlessbox}}

\title{Winds from accretion disks driven by the radiation  and 
magnetocentrifugal force.}

\vspace{1.cm}
\author{ Daniel Proga }
\vspace{1.cm}
\affil{LHEA, GSFC, NASA, Code 662, Greenbelt, MD 20771; proga@sobolev.gsfc.nasa.gov}

\begin{abstract}

We study the two-dimensional, time-dependent hydrodynamics
of  radiation-driven winds  from luminous accretion disks threaded
by a strong, large-scale, ordered magnetic field.
The radiation force is mediated 
primarily by spectral lines and is calculated  using a generalized 
multidimensional formulation of the Sobolev approximation. 
The effects of the magnetic field are approximated by adding
into the equation of motion a force that emulates a magnetocentrifugal force.
Our approach allows us to calculate disk winds when the magnetic
field controls the  geometry of the flow, forces  the flow to corotate 
with the disk, or both. In particular, we calculate models where  
the  lines of the poloidal component of the field are straight and 
inclined to the disk at a fixed angle.

Our numerical calculations show that flows which conserve  specific angular
velocity have a larger mass loss rate  than their
counterparts which conserve specific angular momentum.
The difference in the mass loss rate between these two types of winds 
can be several orders of magnitude for low disk luminosities but 
vanishes for high  disk luminosities. 
Winds which conserve angular velocity 
have  much higher velocities than 
angular momentum conserving winds.
Fixing the wind geometry stabilizes winds which are unsteady
when the geometry is derived self-consistently.
The inclination angle between the poloidal velocity and the normal to the disk 
midplane is important. Non-zero inclination angles allow 
the magnetocentrifugal force to increase
the mass loss rate for low luminosities, and increase the wind velocity
for all luminosities. The presence of the azimuthal force does not 
change the mass loss rate when the geometry of the flow is fixed.

Our calculations also show that the radiation force can launch winds
from  magnetized disks. The line force  can be essential in producing 
magnetohydrodynamical (MHD) winds
from disks where the thermal energy is too low to launch winds
or where the field lines make an angle of $< 30^o$ with respect to the normal
to the disk midplane. In the latter case the wind will be less decollimated
near its base by the centrifugal force and its collimation far away from
the disk can be larger than the collimation of its centrifugally-driven MHD 
counterpart.

\end{abstract}


\section{Introduction}

Accretion disks are believed to lose mass via powerful outflows 
in many astrophysical environments
such as active galactic nuclei (AGN); many types of interacting 
binary stars, non-magnetic cataclysmic variables (nMCVs), for instance; and 
young stellar objects (YSOs).  Magnetically driven
winds from disks are the favored explanation for the outflows
in many of these environments.   
Blandford \& Payne 1982 
(see also Pelletier \& Pudritz 1992)
showed that  the centrifugal force can  drive 
a wind from  the disk if the poloidal component of 
the magnetic field, ${\bf B_p}$ makes an angle of $> 30^o$ with respect to the normal to 
the disk surface. Generally, centrifugally-driven MHD  disk winds 
(magnetocentrifugal 
winds for short) require  the presence of a sufficiently
strong, large-scale, ordered magnetic field threading the disk
and the poloidal magnetic field to be comparable to the toroidal magnetic
field, $|B_\phi/B_p| \simless 1$
(e.g., Cannizzo \& Pudritz 1988, Pelletier \& Pudritz 1992).
Additionally, magnetocentrifugal winds require some thermal 
assistance  to flow freely and steadily from the surface of the 
disk,  to pass through a slow magnetosonic surface 
(e.g., Blandford \& Payne 1982).  

Many authors have  studied  magnetocentrifugal winds from
a Keplerian disk (e.g., Ouyed \& Pudritz 1997, Ustyugova et al. 1999,
Krasnopolsky, Li \& Blandford 1999,
and references therein). These studies are either
analytic, looking for stationary, often self-similar solutions
or numerical, looking for both stationary and time-dependent solutions. 
However in numerical simulations of magnetocentrifugal winds, 
the lower boundary is between
the slow magnetosonic surface  and the  Alfv${\acute{\rm e}}$n surface.
This specification requires setting the mass loss rate in advance
(e.g., Bogovalov 1997).
Simulations of magnetocentrifugal winds  considering the whole disk
(even  regions below the slow magnetosonic surface)
and not requiring an {\it ad hoc} mass 
loss rate, are only now  becoming feasible. 
This requires, however, an accurate 
treatment of the radiation and gas pressure effects, among other physical
processes.

Thermal expansion and the radiation force have been suggested as other
mechanisms that can drive disk winds.
These mechanisms can produce powerful winds without the presence of a 
magnetic field.
Winds are likely thermally driven in 
X-ray-irradiated accretion disks in systems such as  X-ray binaries and
AGNs
(e.g., Begelman, McKee \& Shields 1983, Woods et al. 1996).
These winds require the gas temperature to be so high that
the gas is not gravitationally bound. 
In such cases, the radiation driving
is probably not important because the gas is fully ionized, at least 
in the hottest regions, and the radiation force is only due to 
electron scattering.
However in the  regions where gas is cooler and not fully ionized 
the radiation force will be enhanced by spectral lines and  play an important
role in controlling the dynamics of the flow. In fact Murray et al. (1995)
designed a line-driven disk wind model specifically for AGNs.

Radiation-driven disk winds have been extensively modeled 
(e.g., Pereyra, Kallman \& Blondin 1997; 
Proga, Stone \& Drew 1998, hereafter PSD~I; Proga 1999;
Proga, Stone \& Drew 1999, hereafter PSD~II).
These recent studies showed that radiation pressure due to spectral lines 
can drive winds from luminous disks. This result
has been expected (e.g., Vitello \& Shlosman 1988). 
These studies, in particular those by PSD~I,  
also showed some unexpected 
results. For example, the flow is unsteady in cases where the disk
luminosity dominates the driving radiation field.
Despite the complex structure of the unsteady disk wind,
the time-averaged mass loss rate and terminal velocity scale
with luminosity, as do steady flows obtained where the
radiation is dominated by the central object.

To calculate the line force for disk winds,
PSD~I adopted the method introduced
by Castor, Abbott \& Klein (1975, hereafter CAK) and further
developed by Friend \& Abbott (1986), Pauldrach, Puls \& Kudritzki (1986) 
for one-dimensional radial flows within the context of a  wind from hot, 
luminous OB stars.  To extend the CAK method to model multi-dimensional disk
winds, it is necessary to accommodate the effects of the three-dimensional 
velocity field and the direction-dependent intensity. 
Owocki, Cranmer \& Gayley (1996) showed that 
these effects can lead to qualitatively different results 
compared to those obtained from a one-dimensional treatment
in the case of a rapidly rotating star. 
The most difficult aspect of calculating the line force due to a disk is
in the evaluation of  the integral involving the velocity gradient tensor 
over the entire solid angle occupied by the radiating surface.
PSD~I used an angle-adaptive quadrature to ensure an
accurate result.
However, computational limitations required that they
simplified the integrand, retaining only the dominant terms in the
velocity gradient tensor. 

PSD~II generalized the PSD~I method and introduced a new quadrature that
avoids any simplification of the integrand.  This allowed them to evaluate
the radiation force for completely arbitrary velocity fields within the
context of the CAK formalism.   They applied this method to 
recalculate several disk wind models first discussed in PSD~I.
These more physically accurate models show that PSD~I's more approximate
method was very robust. The PSD~II calculations predict total mass loss
rates and velocities marginally different from those published in PSD~I.
Additionally, PSD~II find that models which display   
unsteady  behavior in PSD~I are also unsteady with the new method.
The largest change
caused by the new method is in the disk-wind opening angle: winds
driven only by the disk radiation are more polar with the new method
while winds driven by the disk and central object radiation
are typically more equatorial.

In this paper, we will numerically  check how strong, ordered magnetic 
fields can change
disk winds driven by the line force for a given disk luminosity. 
We assume that the transport of angular momentum in the disk is dominated 
by local disk viscosity, for instance due to the local shear instability 
in weakly magnetized disk (Balbus \& Hawley 1997). 
Instead of adding the magnetic
fields to the PSD~II model and solving consistently the  equations of MHD, 
we simply start by   adding some of the effects 
due to the magnetic fields, namely (1) the azimuthal force so 
the wind  conserves the specific angular velocity along the streamlines and 
(2) the force perpendicular to the field lines so the geometry of 
the disk wind is controlled by the magnetic field. We thus
adopt the popular concept that the magnetic field dominates outside 
the disk and at  least near the disk surface, one can think of the magnetic
field lines as rigid wires that control any flow (e.g., 
Blandford \& Payne 1982; Pelletier \& Pudritz 1992).
Such an approach is clearly simplistic and can be applied only
to sub-Alfv${\acute{\rm e}}$nic flows.
Nevertheless, the results presented here provide a useful exploratory
study of  line-driven disk winds in the presence of  a strong, large-scale 
field magnetic threading the disk.

The organization of the paper is as follows. 
In Section 2 we describe our numerical calculations; 
in Section 3 we present our results; 
in Section 4 we conclude with a brief discussion. 

\section{Method}

To calculate the structure and evolution of a wind
from a disk, we solve the equations of hydrodynamics
\begin{equation}
   \frac{D\rho}{Dt} + \rho \nabla \cdot {\bf v} = 0,
\end{equation}
\begin{equation}
   \rho \frac{D{\bf v}}{Dt} = - \nabla (\rho c_s^2) + \rho {\bf g}
 + \rho {\bf F}^{rad} + \rho {\bf F}^{mc}
\end{equation}
where $\rho$ is the mass density, ${\bf v}$ the velocity,
${\bf g}$ the gravitational acceleration of the central object,
${\bf F}^{rad}$ the total radiation force per unit mass, and
${\bf F}^{mc}$ the total 'magnetocentrifugal' force. 
The term with ${\bf F}^{mc}$ 
ensures that  gas flows
along an assumed direction  and conserves its specific angular velocity.
The gas in the wind is  isothermal with a sound speed $c_s$.
We solve these equations in spherical polar coordinates $(\rho,\theta,\phi)$.

The geometry and assumptions needed to compute the radiation
field from the disk and central object are as in PSD~II (see also PSD~I).
The disk is flat, Keplerian, geometrically-thin and
optically-thick.  
We specify the radiation field of the disk  by
assuming that the temperature follows the radial profile of the so-called
$\alpha$-disk (Shakura \& Sunyaev 1973), and therefore depends only on
the mass accretion rate in the disk, $\dot{M}_a$,  and the mass 
and radius of the central object, $M_\ast$  and $r_\ast$.  In particular,
the disk luminosity, $L_D \equiv GM_\ast \MDOT_a/2r_\ast$.
In models where the central object  radiates, we take into account 
the stellar irradiation of the disk, assuming that the disk re-emits all 
absorbed energy locally and isotropically. We express the central object 
luminosity $L_\ast$ in units of the disk luminosity $L_\ast=x L_D$.
See PSD~I and PSD~II for further details.

Our numerical algorithm
for evaluating the line force is described in PSD~II.  
For a rotating flow, there may be an azimuthal component to the line force 
even in axisymmetry. However we set this component of the
line force to zero because it is rather weak as compared to other components
and is not of great importance (e.g., PSD~II). 
Then we assume that the rotational velocity $v_\phi$ is
determined by the azimuthal component of the magnetocentrifugal force. 
The description of our calculation of the 'magnetocentrifugal' force
follows.

We choose the simplest geometry of the magnetic field and the flow: 
the poloidal component of the magnetic field, $\bf B_p$ and 
of the velocity, ${\bf v}_p$ are 
parallel to one another, and 
the inclination angle $i$ between ${\bf v}_p$
and the disk midplane
is  fixed for all locations and times. 
In other words, we constrain the geometry of the flow to straight
cones of pre-specified, radius- and time-independent opening angle.
In practice, we impose this geometry in the following way: 
(i) we evaluate the physical accelerations and the curvature terms 
in  eq. 2, hereafter collectively referred to as   
\begin{equation}
{\bf F}'    = - \frac{1}{\rho} \nabla (\rho c_s^2) + {\bf g}
 + {\bf F}^{rad}    + {\bf F}^{i}, 
\end{equation}
where ${\bf F}^{i}$ represents the curvature terms appearing on the
left hand side of eq. 2 when $\frac{D{\bf v}}{Dt}$ is expressed in the
spherical polar coordinate system (e.g., Shu 1992):
\begin{equation}
{\bf F}^i=  
             \left ( \begin{array}{c} 
                    \frac{v_\theta^2+v_\phi^2}{r}\\
                    \frac{v_\phi^2}{r} \cot{\theta}\\
                                   0 \\
                      \end{array} \right),
\end{equation}
(ii) we calculate the component of $\bf F'$ 
perpendicular to a streamline:
\begin{equation}
{\bf F'}_\perp = {\bf \Lambda} {\bf F'}, 
\end{equation}
where
\begin{equation}
{\bf {\Lambda}} =
\left ( \begin{array}{ccc} 
          \sin^2{(\theta -i)} & 
\sin{(\theta -i)}\cos{(\theta -i)} & 0 \\
          \sin{(\theta -i)}\cos{(\theta -i)} 
& \cos^2{(\theta -i)}      & 0 \\
          0              & 0              & 0 
        \end{array} \right),
\end{equation}
(iii) and finally we subtract $\bf F'_\perp$ from $\bf F'$ so 
the total remaining force is tangent to the streamline. 

The disk outflow that conserves its specific angular velocity,
$\Omega=v_\phi/ (r \sin\theta)$ along a streamline is simply the flow
that corotates with the disk, i.e., $\Omega$ is constant and
equals to the disk rotational velocity, $\Omega_D=v_\phi(r_D, 90^o)/r_D$ 
at the footpoint of the streamline
on the disk at the radius, $r_D$.
To ensure corotation of the wind, we introduce the azimuthal force:
\begin{equation}
{\bf F}^t=  
             \left ( \begin{array}{c} 
                                   0 \\
                                   0 \\
     2\frac{ v_\phi (v_r \sin{\theta} +v_\theta \cos{\theta})} {r \sin{\theta}}  \\
                      \end{array} \right).
\end{equation}

Summarizing, our magnetocentrifugal force is
\begin{equation}
{\bf F}^{mc} = -{\bf \Lambda} {\bf F'} + {\bf F}^t. 
\end{equation}
As in PSD~II, we use the ZEUS-2D code (Stone \& Norman 1992)  
to numerically integrate the hydrodynamical equations (1) and (2).
We did not change the transport step in the code and so the changes we made
are just to the source step as outlined above.

We explore  disk wind models for three cases:
(I) a wind corotates with the disk 
(${\bf F'}_\perp =0$ and ${F_\phi}^t = 2 v_\phi (v_r \sin{\theta} +v_\theta \cos{\theta})/(r \sin{\theta} )$) ; 
(II) the wind geometry is fixed
by adopting a fixed  inclination angle between the poloidal component of 
the velocity and the disk midplane 
( ${\bf F'}_\perp = \bf{ \Lambda F'}$ and ${\bf F}^t = 0$) ; and 
(III) a combination of (I) and (II) ( ${\bf F'}_\perp = \bf{ \Lambda F'}$ and 
${F_\phi}^t =  2 v_\phi (v_r \sin{\theta} +v_\theta \cos{\theta})/(r \sin{\theta}) $). 

In case I with $x=0$, we decided to put an additional constraint on the radial
velocity to produce outward flows.
Near the disk midplane, fluctuations 
of the density and velocity can occur (e.g., PSD~I). 
Additionally, the radial component of the radiation force is negative
near the disk midplane and the central object (e.g., Icke 1980, PSD~I).
Thus $v_r$ can be less than zero;  gravity can dominate
and  pull the flow toward the center. To avoid this   
we apply the condition:  $v_r={\rm max}(v_r, -\cot \theta v_\theta)$.
This condition does not change the mass loss rate, but does  reduce the 
wind opening angle.

To produce transonic flows in cases~II and III, we need to increase 
the density along the disk midplane, $\rho_o$, 
from $10^{-9}~{\rm g~cm^{-3}}$  (as used by PSD~I)
up to $10^{-4} ~{\rm g~cm^{-3}}$. For low $\rho_o$, the wind velocity 
does not depend on $\rho$,
the mass loss rate is proportional to $\rho_o$ and may be higher
than the mass accretion rate. For $\rho_o \simgreat
10^{-4} ~{\rm g~cm^{-3}}$ the mass loss rate is dramatically lower,
does not change with $\rho_o$,
and the flow is subsonic near the disk midplane.
These are  desired properties of the outflow; we expect  gas near 
the midplane
to be nearly in hydrostatic equilibrium. Equally important, we assume
that the disk is in a steady state so the mass loss rate should be
lower than the mass accretion rate. 

In reality, gas near the disk midplane (inside the disk) 
is in hydrostatic equilibrium 
for the densities lower than those we assumed 
because the radiation force is lower as  the 
radiation becomes  isotropic.  
We would like to stress that we treat the region very close to 
the midplane as a boundary condition
and do not attempt to model the disk interior. 

As in PSD~I and PSD~II, 
we calculate disk winds with  model parameters suitable for a typical 
nMCV (see PSD~I's  Table~1 and our Table~1). We vary the disk and central
object luminosity and the inclination angle. We hold  all other
parameters fixed, in particular the parameters
of the CAK force multiplier: $k=0.2$, $\alpha=0.6$ and $M_{max}=4400$
(see PSD~I).
Nevertheless we can use our results to predict the wind properties for
other parameters and  systems -- such as AGN -- by applying 
the dimensionless form of 
the hydrodynamic equations and the scaling formulae as discussed 
in PSD~I and Proga (1999).

\section{Results}

PSD~I and PSD~II  showed that radiation-driven
winds from a disk fall into two categories: 
1) intrinsically unsteady with large
fluctuations in density and velocity, and 2) steady with smooth density 
and velocity. The type of the flow is set by the geometry of the
radiation field, parametrized by $x$: if the radiation field is dominated
by the disk ($x<1$) then the flow is unsteady, and if the radiation is
dominated by the central object ($x\simgreat 1$) then the flow is steady.
The geometry of the radiation field also determines the geometry of the
flow; the wind becomes more polar as $x$ decreases. However the mass-loss
rate and terminal velocity are insensitive to geometry and depend more
on the total system luminosity, $L_D+L_\ast$. We recalculated some
of the PSD~II models to check how inclusion of the magnetocentrifugal
force will change line-driven disk winds.

Figure~1 compares the density in the wind in two models from PSD~II 
where they used a generalized CAK method (top panels), models using
PSD~II's method but conserving specific angular velocity -- our case~I 
(middle panels), and models using PSD~II's method, conserving specific angular 
velocity and having fixed inclination angle of the stream lines --
our case~III (bottom panels). Models in case~III (bottom panels) are with
the inclination angles approximately equal to those in the PSD~II
case (the top panels).
The left column shows the results for a model with 
$\dot{M}_a= 10^{-8}~\MSUNYR$ and $x=0$, while the right column shows 
the results with $\dot{M}_a= \pi \times 10^{-8}~\MSUNYR$ and $x=1$.  
The former corresponds to the fiducial $x=0$ model while the latter 
corresponds to the fiducial $x=1$ model discussed in detail in PSD~I 
and PSD~II.

We start with describing  the $x=0$ wind model 
(left column panels). 
The most striking yet expected difference between PSD~II's case (top) 
and our  case I  (middle) is a decrease of the wind opening angle, $\omega$ 
from $50^o$ to $15^o$ when the specific angular momentum conservation 
is replaced by the specific angular velocity conservation. 
This result is  caused by a much stronger
centrifugal force in our case I than in the PSD~II case which is also reflected
in an increase of the radial velocity by more than one order of magnitude
(Table~1). Another big difference between  the two cases
is in the mass loss rate, $\dot{M}_w$ which increases from
$5.5\times 10^{-14}~\MSUNYR$ to $1.3\times 10^{-12}~\MSUNYR$. 
An unchanged property in these
two cases is that the wind is unsteady.
The quantitative changes between PSD~I's case (top) and case III (bottom)
are similar or smaller than the changes described above (see Table~1).
The most pronounced change caused by 
holding the inclination angle fixed is that the wind becomes steady.
However this is not surprising because models with the fixed, rigid geometry 
are pseudo one-dimensional and the flow has one degree of freedom in space.
In PSD~I and PSD~II models where the wind geometry is calculated
self-consistently, a strong radial radiation force is required 
to produce a steady  outflow.

The x=1 wind model (top right hand panel of Figure~1) is an example of 
a steady outflow from PSD~II. This model remains steady 
in the  two other cases shown in the figure
and in   case I listed in  Table~1 (run Ic).
The model changes from case to case mainly in
the radial and rotational velocities which increase
when the azimuthal force is added (middle and bottom panel).
For example, the radial velocity  increased from 3500 ${\rm km~s^{-1}}$
in  PSD~II's run~C (top panel) to 28000 ${\rm km~s^{-1}}$ in our run IIIc
(bottom panel). 
As in the x=0 wind model, adding the azimuthal force also reduces the wind
opening angle (middle panel). However an increase in the mass loss rate
from PSD~II's case (top panel) to our case~II 
(bottom panel) is only a factor of $\sim 2$. 
The mass loss rate is practically
the same in case II and case III (run IIc and IIIc).

Models presented in Figure~1 both for $x=0$ and $x=1$ suggest that
an increase in the mass loss rate due to conservation of the specific angular
velocity decreases with the disk luminosity. Our other results, 
shown in Table~1, confirm this. 
For example, the x=0 model Ib
with $\dot{M}_a=\pi \times 10^{-8}~\MSUNYR$
has $\dot{M}_w$ higher than the corresponding PSD~II's model B by a factor of 
$\sim 3$ while an increase of $\MDOT_w$ for models with 
$\MDOT_a=10^{-8}~\MSUNYR$ is by a factor $\sim 24$ (i.e., runs A and Ia).

The  mass loss rate of the line-driven disk wind  also  increases by 
imposing the wind geometry such that $i> 0^o$ as comparison of our results
in case II and from  PSD~II reveals. Once the geometry is fixed the mass 
loss rate does not change with adding the azimuthal force --  
the corresponding models in our case II and III have the same $\MDOT_w$.

Purely line-driven x=0 winds have streamlines perpendicular to the
disk midplane near the midplane because the total horizontal force
is negligible in comparison with the vertical force due to lines.
This property of  disk winds has been 
assumed in analytic studies  of line-driven disk winds 
(e.g., Vitello \& Shlosman  1988).
We calculated a model assuming $i=0^o$ (run IIa) and find
that the mass loss rate is very similar to the corresponding PSD~II's run~A.
This confirms then that the mass loss rate in line-driven disk winds for $x=0$
is determined in the regions 
where the matter flows still perpendicularly to the disk.

\section{Conclusions and Discussion}

We have studied winds from accretion disks with very strong organized
magnetic fields and the radiation force due to lines.
We use numerical methods to solve the two-dimensional,
time-dependent equations of hydrodynamics. We have accounted for
the radiation force using a generalized multidimensional formulation
of the Sobolev approximation. 
To include the effects of the strong
magnetic field we have added to the hydrodynamic equation an approximate
magnetocentrifugal force. Our approach of treating the latter 
allows us to study cases where the wind geometry is controlled entirely
by the magnetic field with the lines of the poloidal component of the field,
${\bf B}_p$  
being  straight and inclined at a fixed angle to the disk midplane 
where the specific angular velocity is conserved along the streamlines due to 
the magnetic azimuthal force, or both. 

Our simulations of line-driven disk winds show that inclusion 
of the azimuthal force, ${\bf F}^t$ which makes the wind  corotate with 
the disk, increases the wind mass loss rate significantly only 
for low disk luminosities. 
As expected such winds have much higher velocities than winds 
with zero azimuthal force.
Fixing the wind geometry effectively reduces
the flow from two-dimensional to one-dimensional. This in turn 
stabilizes  winds   which are
unsteady when the geometry is allowed to be determined self-consistently.
The inclination angle between the poloidal velocity and the normal to the disk 
midplane is important. If it is higher than zero it can significantly
increase  
the mass loss rate for low luminosities, and increase the wind velocity
for all luminosities. The presence of the azimuthal force does not 
change the mass loss rate if the geometry is fixed.

It is very intriguing that the mass loss rate can be enhanced by the 
magnetocentrifugal force the most for low luminosities where 
line force is  weak and at the same time
the mass loss rate remains a strong function of the disk
luminosity for all luminosities. In particular, there is no
wind when $\dot{M}_a \simless 10^{-8}~\MSUNYR$ as in purely line-driven case
(e.g.,  PSD~I; Proga 1999).

Model IIa' with $i=0$ is steady and has the same $\dot{M_w}$ as  
the corresponding PSD~II model. 
The mass loss rate is then the same regardless of the time behavior. 
This is consistent with PSD~I conclusion that the mass loss rate
depends predominately on the total system luminosity.

Drew \& Proga (1999) applied results from PSD~I, PSD~II and Proga~1999 
to nMCV. In particular, 
they compared mass loss rates predicted by the  models with observational 
constraints. They concluded that either mass accretion rates in high-state
nMCV are higher than presently thought  by a factor of 2-3  or that radiation
pressure alone is not quite sufficient to drive the observed hypersonic 
flows. If the latter were true then 
an obvious candidate to assist radiation pressure in these cases is MHD
(e.g., Drew \& Proga 1999).

An increases of the mass loss rate due to inclusion of 
a magnetocentrifugal force (our case I) brings our predictions close to 
the observational estimates for nMCVs. However we should bear in mind that at 
the same time, the radial velocity of the wind increases above 
the observed velocities in nMCVs that are of the order a few thousand 
${\rm km~s^{-1}}$. 
The increase of the wind velocity may not be  large 
if the magnetic field is moderately strong and spins up the disk wind
only close to its base where the mass loss rate is determined. Then
farther away from the disk the wind would  gain less or no angular momentum 
and velocities would be similar to those in the case without the magnetic 
field at all. 

Outflows which conserve  specific angular velocity have a higher rotational
velocity than those which conserve  specific angular momentum.
The former also have  higher terminal velocities due to stronger
centrifugal force. Additionally in the wind where the specific angular velocity
is conserved the rotational velocity is comparable to the terminal 
velocity
while in the wind where the specific angular momentum is conserved 
the rotational velocity decreases asymptotically to zero with 
increasing radius
and is therefore much lower than the terminal velocity.
Thus we should be able to distinguish these two kinds of winds
based on their line profiles. 
For example, highly rotating  winds should produce emission lines  much 
broader than slowly rotating, expanding winds, if we see
the disk edge-on. 

So far we have discussed changes of the line-driven wind caused
by inclusion of the magnetocentrifugal force. However we can
also anticipate some changes in the magnetocentrifugal wind if we add the line force.
For example, there may be  a difference in the wind collimation caused by 
the line force.
The centrifugal force decollimates the magnetocentrifugal 
outflow near the disk as $i$ must be greater than $30^o$. In such  a case 
the magnetic field collimates the outlow only beyond 
the Alfv${\acute{\rm e}}$n surface where the pinch  force exerted by 
the toroidal component of the field operates.
In the case with a significant line driving, the magnetic field
does not have to be inclined at the angle $> 30^o$. Then in such case, 
outflows are more collimated than pure magnetocentrifugal outflows 
from the start and may end up more collimated far away from the disk.

Let us now discuss some results of  other related works. 
Recently Ogilvie \& Livio (1998) studied a thin
accretion disk threaded by a poloidal magnetic field. Their
purpose was to determine qualitatively how much thermal assistance 
is required for the flow to pass through the slow magnetosonic 
surface.  They found that 
a certain potential difference must be overcome even when $i > 30^o$
and that thermal energy is not sufficient to
launch an outlow from a magnetized disk. 
Ogilvie \& Livio suggested that an additional source of 
energy, such as coronal heating may be required. PSD~I showed
the radiative line force can  drive disk winds. Thus the 'missing'
energy  may be in the radiation field.
Our calculations here show that if the disk luminosity is
sufficiently high the line and magnetocentrifugal forces
produce strong transonic winds regardless of the wind geometry.
Then the magnetocentrifugal force can assist
the line force in producing disk winds -- our approach to the problem 
or the line force or/and the thermal force 
can well assist  MHD -- Ogilvie \& Livio's approach
(see also Wardle \& K$\ddot{\rm o}$ngil 1993;
Cao \& Spruit 1994, for instance).
In seeking a steady state
solution of a line-driven disk wind, Vitello \& Shlosman (1988) also found 
that the flow must overcome a potential difference -- an increase of the 
vertical gravity component with height. They suggested
that for the radiation force to increase with height above the disk 
midplane and overcome this potential difference 
a very particular variation in the ionization state of the 
gas is required. 

There are a number of limitations of our treatment of magnetic fields
which are worthy of mention.

Our models  do not include the whole richness of MHD 
because we approximate the Lorentz force by
the 'magnetocentrifugal' force (eqs 3-8) instead of
solving self-consistently
the equation of motion and the induction equation. 
We have included effects of a magnetic field on the disk winds 
in a  simplistic manner that mimics some effects of a very strong, 
organized magnetic field where $B_\phi \simless B_p$ near the disk surface. 
The magnetic field is treated as a rigid
wire that controls the flow  geometry outside the disk. More quantitatively, 
this corresponds to the situation where outside the
disk, at least  near the disk photosphere, the magnetic field dominates i.e., 
the magnetic pressure  exceeds the disk gas pressure  
$\beta\equiv 8\pi p_D/ B^2_D < 1 $.
Inside the disk however the situation may be different.
We assume that the disk is in a steady state that is also stable.
In such a case, the regions near the disk midplane 
are likely supported by the gas pressure and 
the magnetic pressure should be  smaller than the gas pressure, 
i.e., $\beta   \simgreat 1$, otherwise the disk may be unstable 
(Stella \& Rosner 1984). 
Using the system parameters adopted here: $c_s= 14~\rm km~s^{-1}$ and
$\rho_o=10^{-4}~\rm g~cm^{-3}$, we find that the $\beta \simgreat 1$ 
condition yields a maximum value for the disk magnetic fields strentgh 
of the order of $\sim~10^5$~G.
Our assumption that poloidal magnetic field lines are straight excludes
any magnetic tension in the $(r,\theta)$ plane. 

In our cases~I and III we assume that the specific angular velocity, 
$\Omega$ is conserved  along a streamline from the footpoint of the line  
to infinity or rather to  the outer boundary of the computational domain. 
However in a steady state axisymmetric MHD flow, the quantity which is 
conserved is the total angular momentum per unit mass which can be written as 
\begin{equation}
l=\Omega r^2 \sin^2\theta - \frac{r \sin\theta B_\phi} {4 \pi\kappa}
\equiv \Omega_D r_A^2,
\end{equation}
where $r_A$ is the position of the the Alfv${\acute{\rm e}}$n point of 
the flow on the streamline, and $\kappa=\rho v_p/B_p$ is the mass load--
another constant along the streamline (e.g., Pelletier \& Pudritz 1992). 
The total angular momentum has contributions from both the flowing
rotating gas and the twisted magnetic field. This means that 
as the material
angular momentum, $\Omega r^2 \sin^2\theta $ 
increases due to corotation 
with the disk, the toroidal
component of the magnetic field must increase so the total
angular momentum is conserved.
Then our approximation in cases~I and III corresponds to the situation where 
near the disk the total angular momentum is dominated by the contribution
from the twisted magnetic field and  the Alfv${\acute{\rm e}}$nic
surface is beyond our computational domain. In other words our approximation
is valid in the region where the outflow is sub-Alfv${\acute{\rm e}}$nic.  
Consequently,  our approach  does not allow us to study collimation of 
the wind by the magnetic field because this happens beyond 
the fast magnetosonic surface where the flow 
is super-Alfv${\acute{\rm e}}$nic.

PSD~II's models  and ours in case II correspond to 
the other extreme case where the total angular momentum 
has no contribution from the twisted magnetic field and the
material angular momentum is conserved. 

Our treatment of magnetic fields does not include effects of 
the magnetic pressure. 
The magnetocentrifugal driving presupposes the existence of a strong
poloidal magnetic field comparable to the toroidal magnetic field near 
the disk surface,  $|B_\phi/B_p| \simless 1$.  However when 
the poloidal magnetic field is weaker than the toroidal magnetic field, 
$|B_\phi/B_p| >> 1$ 
the magnetic pressure may be dynamically important in driving disk winds
(e.g., Uchida \& Shibata 1985; 
Pudritz \& Norman 1986, Shibata \& Uchida 1986,
Contopoulos 1995; Kudoh \& Shibata 1997; 
Ouyed \& Pudritz 1997 and references therein).
For $|B_\phi/ B_p| >> 1$, there is initially the buildup of 
the toroidal magnetic field by the differential rotation of the disk 
that in turn generates the
magnetic pressure of the toroidal field. The magnetic pressure then
gives rise to a self-starting wind. To produce a steady outflow  
driven by the magnetic pressure a steady supply of the advected
toroidal magnetic flux at the wind base is needed, otherwise
the outflow is likely a transient (e.g., K$\ddot{\rm o}$nigl 1993,
Contopoulos 1995, Ouyed \& Pudritz 1997). However it is not clear
whether the differential rotation of the disk can produce such a supply
of the toroidal magnetic flux to match the escape of magnetic flux
in the wind and even if it does whether such a system will be stable  
(e.g., Contopoulos 1995, Ouyed \& Pudritz 1997 and references therein).

Concluding, we would like to stress that 
a further development of models of radiation-driven winds from disks should 
include taking into account magnetic field but it is equally important to 
consider adding the radiation force to models of MHD winds from luminous disks.
Both kinds of models are quite well understood now and if merged they could 
allow us to study better disk winds in systems such as cataclysmic
variables and AGNs.
As we mentioned above thermal assistance may not be sufficient
to launch wind from a magnetized disk and we should consider not only
coronal heating but also line driving. 

ACKNOWLEDGEMENTS: We thank John Cannizzo, Janet Drew, Achim Feldmeier, 
Tim Kallman, Scott Kenyon, Mario Livio, and James Stone 
for comments on earlier drafts of this paper. 
We also thank Steven Shore and an anonymous referee for comments
that helped us clarify our presentation.
This work was performed while the author held
a National Research Council Research Associateship at NASA/GSFC.
Computations were supported by NASA grant NRA-97-12-055-154.

\newpage
\section*{ REFERENCES}
 \everypar=
   {\hangafter=1 \hangindent=.5in}

{

  Balbus, S.A., \& Hawley, J.F. 1998, Rev. Mod. Phys., 70, 1 

  Begelman M.C., McKee C.F., Shields G. A. 1983, ApJ, 271, 70

  Blandford R.D., Payne D.G. 1982, MNRAS, 199, 883

  Bogovalov S.V. 1997, A\&A, 323, 634

  Cannizzo J. K., Pudritz R.E. 1988, ApJ, 327, 840

  Cao X., Spruit H.C. 1994, A\&A, 287, 80

  Castor J.I., Abbott D.C.,  Klein R.I. 1975, ApJ, 195, 157 (CAK)
  
  Contopoulos J. 1995, ApJ, 450, 616

  Drew J.E., Proga D., in  {\it  ``Cataclysmic Variables'',
  Symposium in Honour of Brian Warner}, Oxford 1999, 
  ed. by P. Charles,  A. King, D. O'Donoghue, in press   

  Friend D.B., Abbott D.C., 1986 ApJ, 311, 701

  Icke V. 1980, AJ, 85, 329

  K$\ddot{\rm o}$nigl A. 1993, in  {\it ``Astrophysical Jets''},
  ed. by D.P. O'Dea (Cambridge: Cambridge Univ. Press), 239
  
  Krasnopolsky R., Li Z.-Y., Blandford R., 1999, ApJ, 526, 631 

  Kudoh T., Shibata K. 1997, ApJ, 474, 362

  Murray N., Chiang J., Grossman S.A., Voit G.M. 1995, ApJ, 451, 498
 
  Ogilvie G.I., Livio M. 1998, ApJ, 499, 329
  
  Ouyed R., Pudritz R.E. 1997, ApJ, 484, 794

  Owocki S.P., Cranmer S.R.,  Gayley K.G.  1996, ApJ, 472, L115 

  Pauldrach A., Puls J.,  Kudritzki R.P. 1986, A\&A, 164, 86 

  Pelletier G., Pudritz R.E. 1992, ApJ, 394, 117

  Pereyra N.A., Kallman T.R. Blondin J.M. 1997, ApJ, 477, 368

  Proga D. 1999, MNRAS, 304, 938

  Proga D., Stone J.M., Drew J.E. 1998, MNRAS, 295, 595 (PSD~I)

  Proga D., Stone J.M., Drew J.E. 1999, MNRAS, 310, 476  (PSD~II)

  Pudritz R.E., Norman C.A. 1986, ApJ, 301, 571 

  Shakura N.I., Sunyaev R.A. 1973 A\&A, 24, 337

  Shibata K.,  Uchida Y. 1986, PASJ, 38, 631
 
  Shu, F. 1992, The Physics of Astrophysics, Vol. 2, Gas dynamics 
     (Mill Valley: University Science Books)
  
  Stellar L., Rosner R. 1984, ApJ, 277, 312
  
  Stone J.M., Norman M.L. 1994, ApJ, 433, 746

  Uchida Y., Shibata K. 1985, PASJ, 37, 515

  Ustyugova G.V., Koldoba A.V., Romanova M.M. Chechetkin V.M. Lovelace R.V.E.
  1999, ApJ, 516, 221

  Vitello P.A.J.,  Shlosman I., 1988 ApJ, 327, 680

  Wardle M., K$\ddot{\rm o}$nigl A. 1993, ApJ, 410, 218

  Woods, D. T., Klein, R. I., Castor, J. I., McKee, C. F., Bell, J. B.
  1996, ApJ, 461, 767
}

\newpage

\begin{figure}
\begin{picture}(280,590)
\put(140,0){\includegraphics{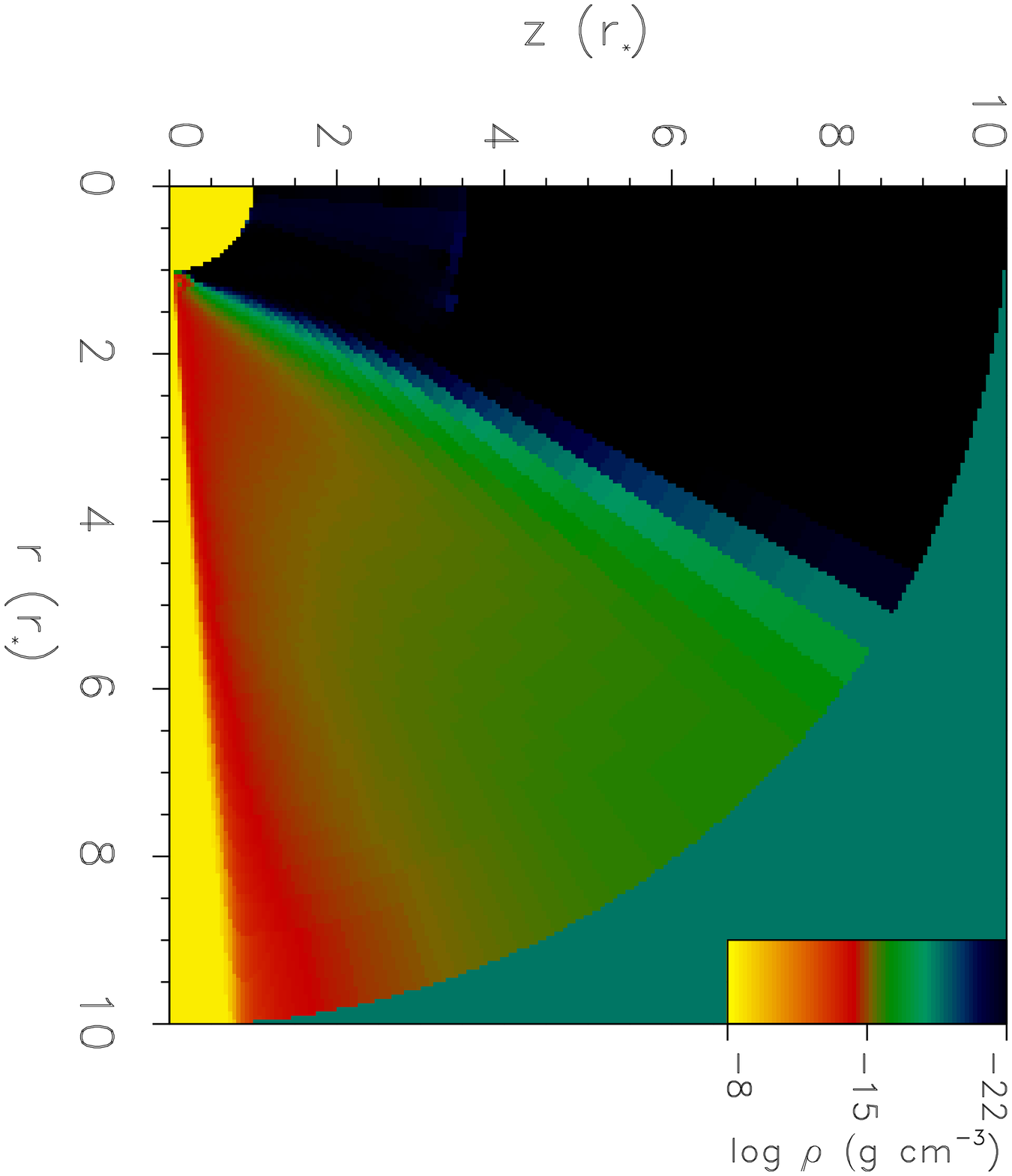}}
\put(140,205){\includegraphics{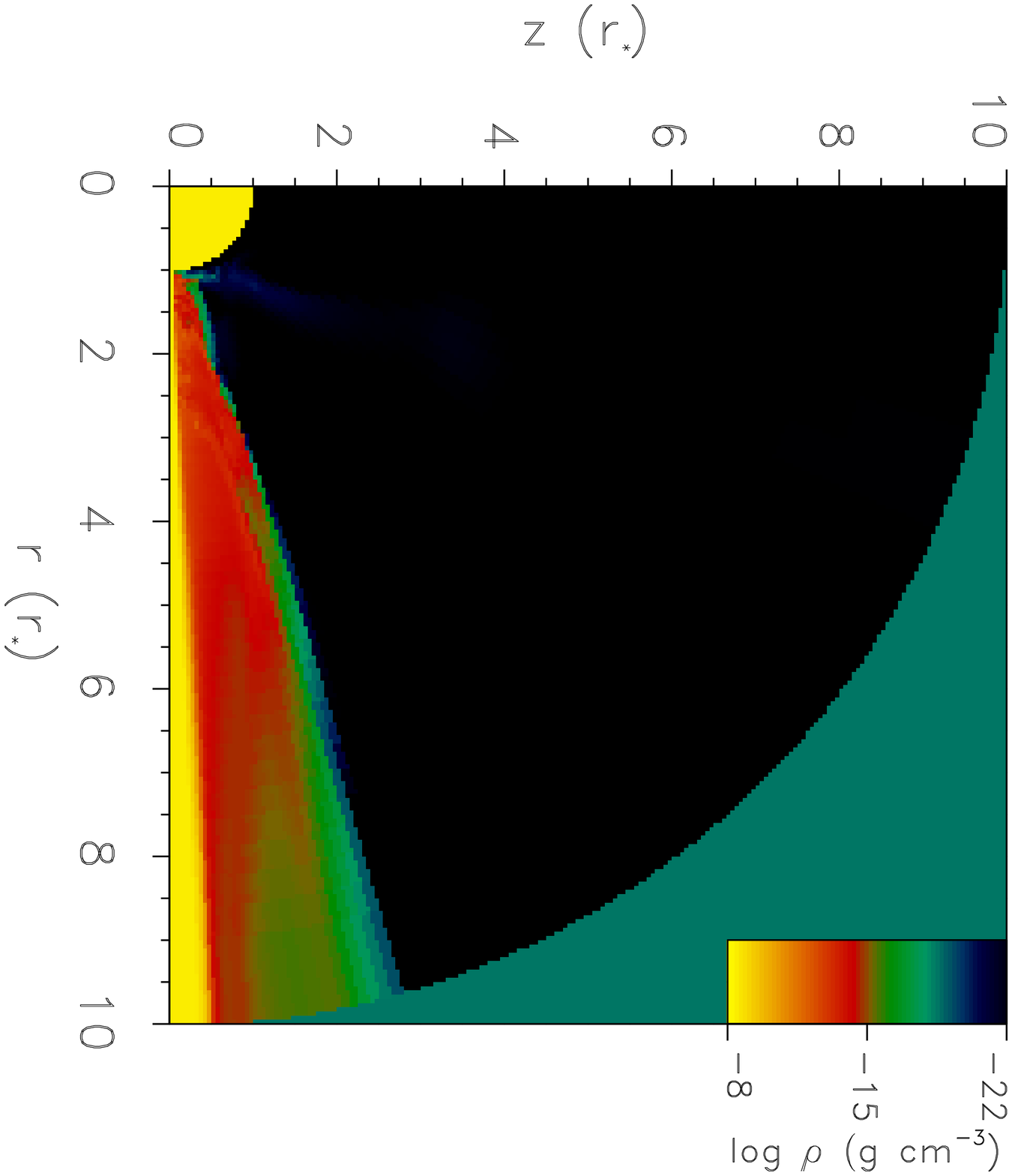}}
\put(140,410){\includegraphics{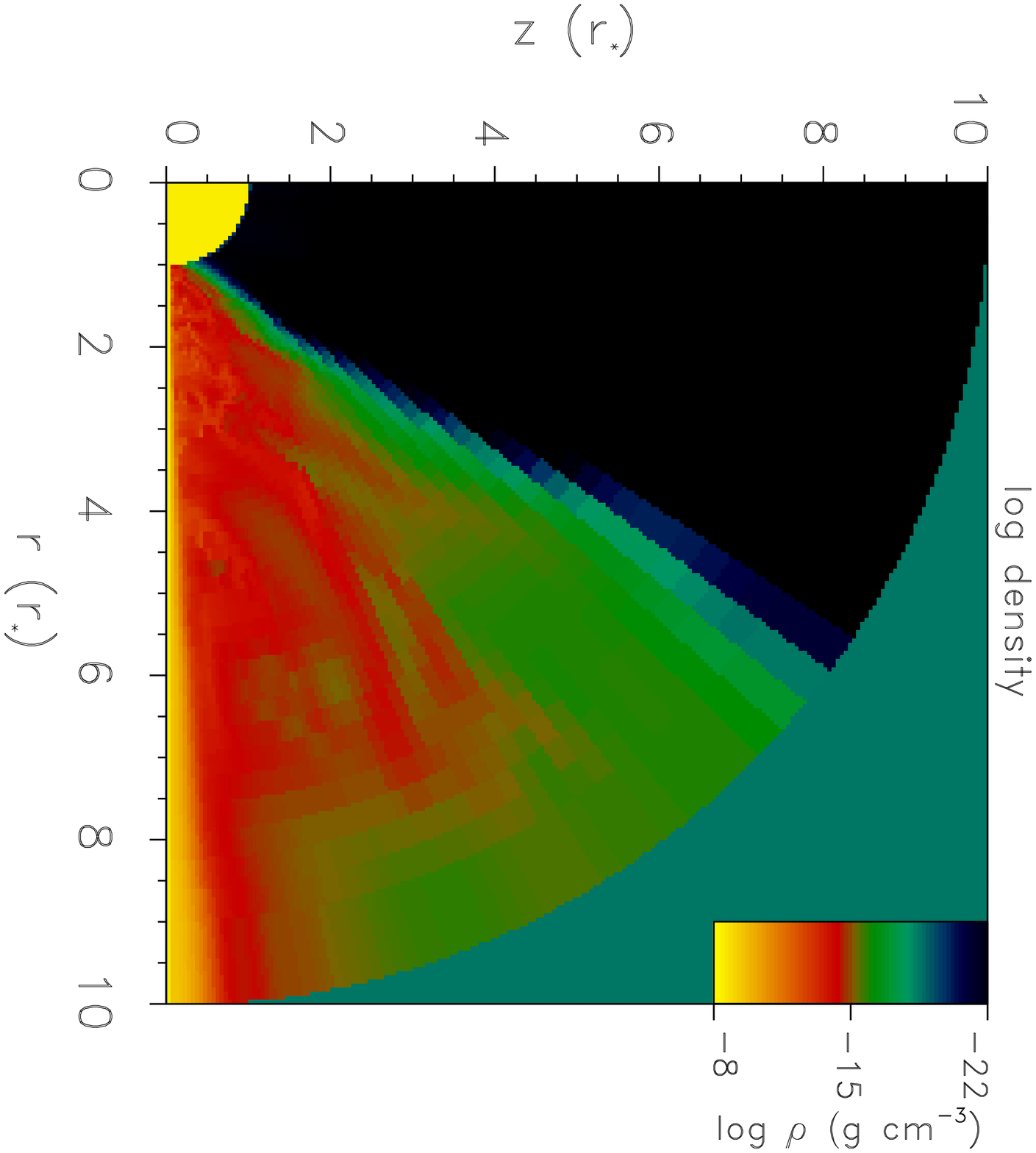}}
\put(230,0){\includegraphics{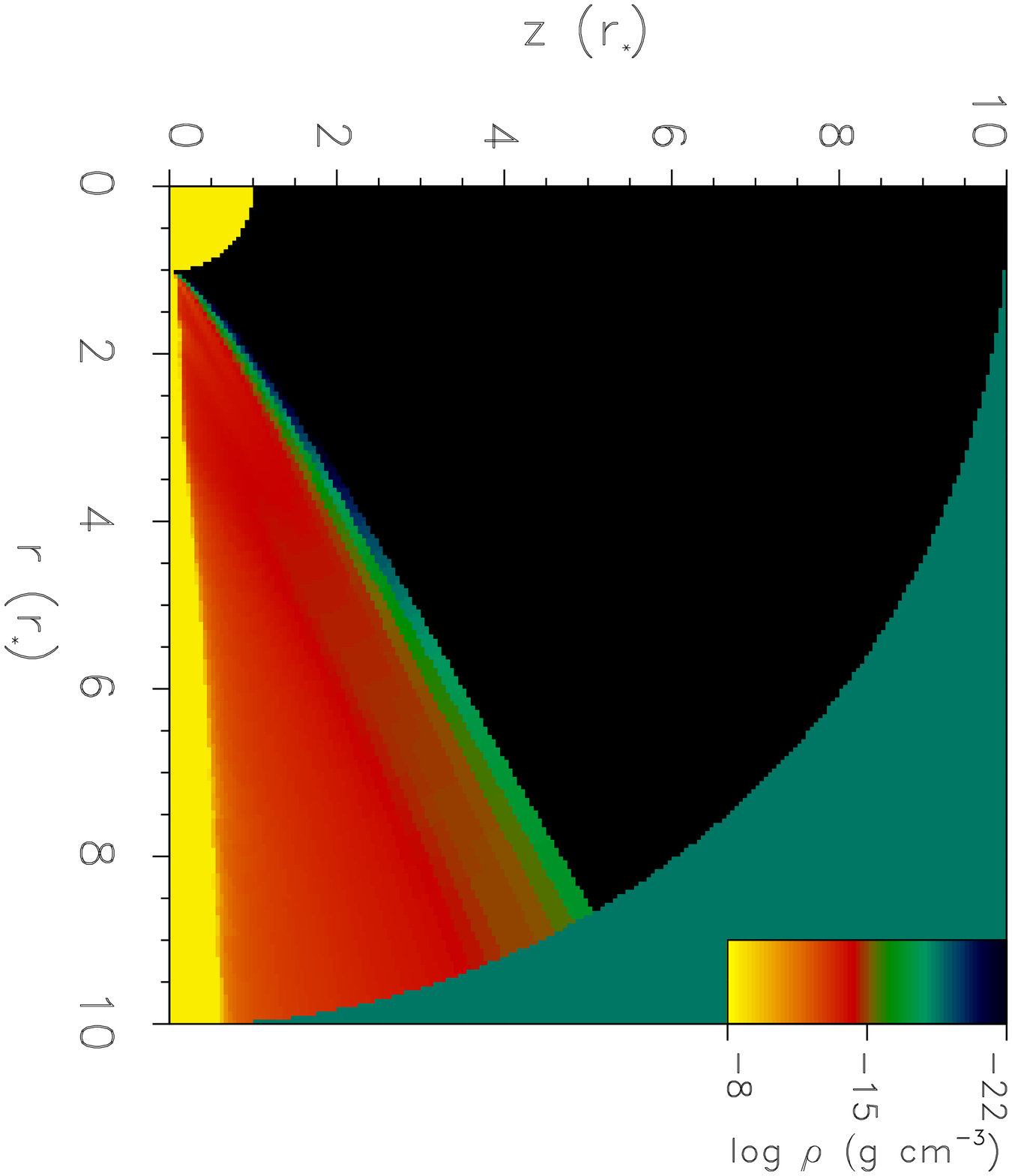}}
\put(230,205){\includegraphics{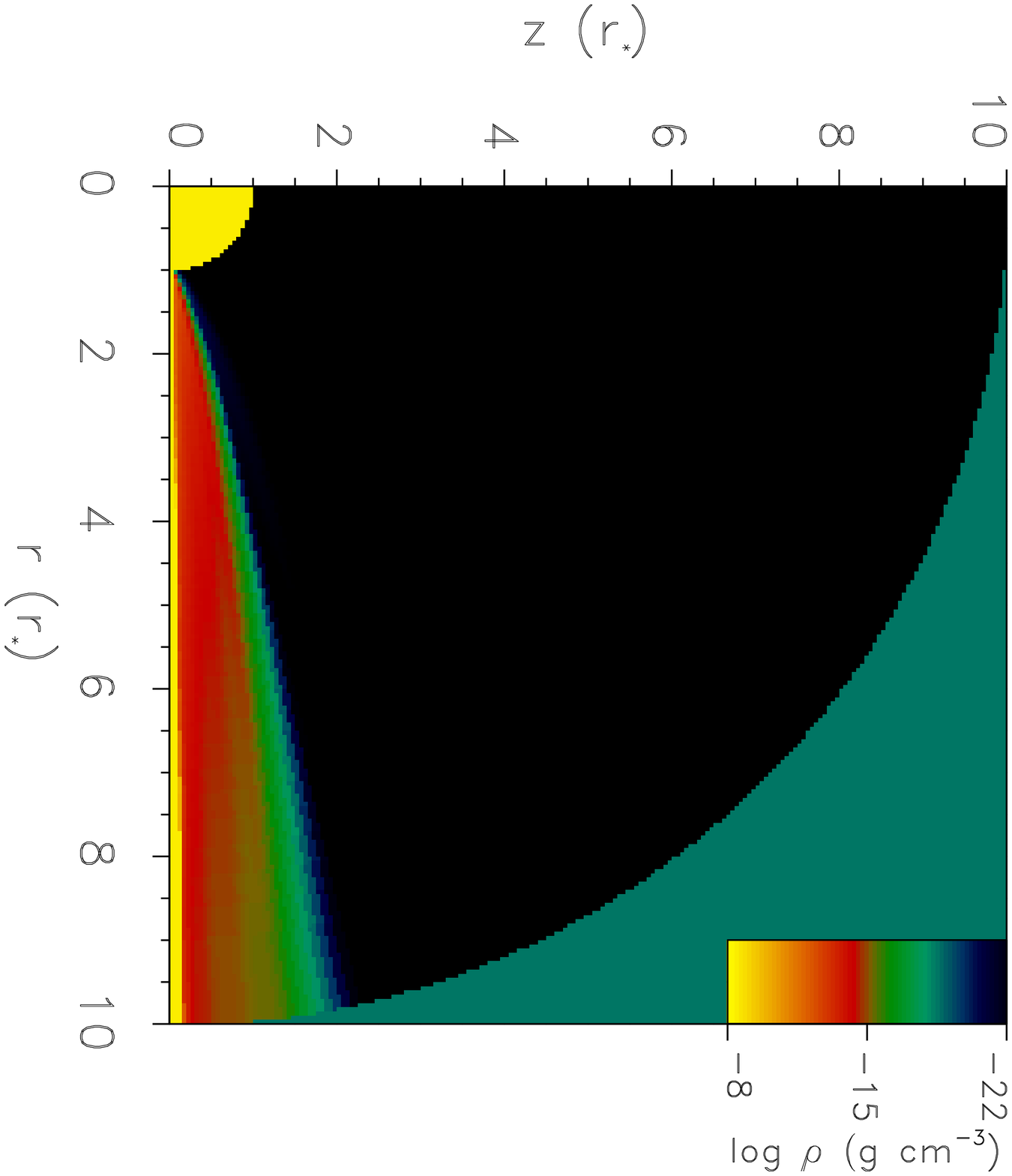}}
\put(230,410){\includegraphics{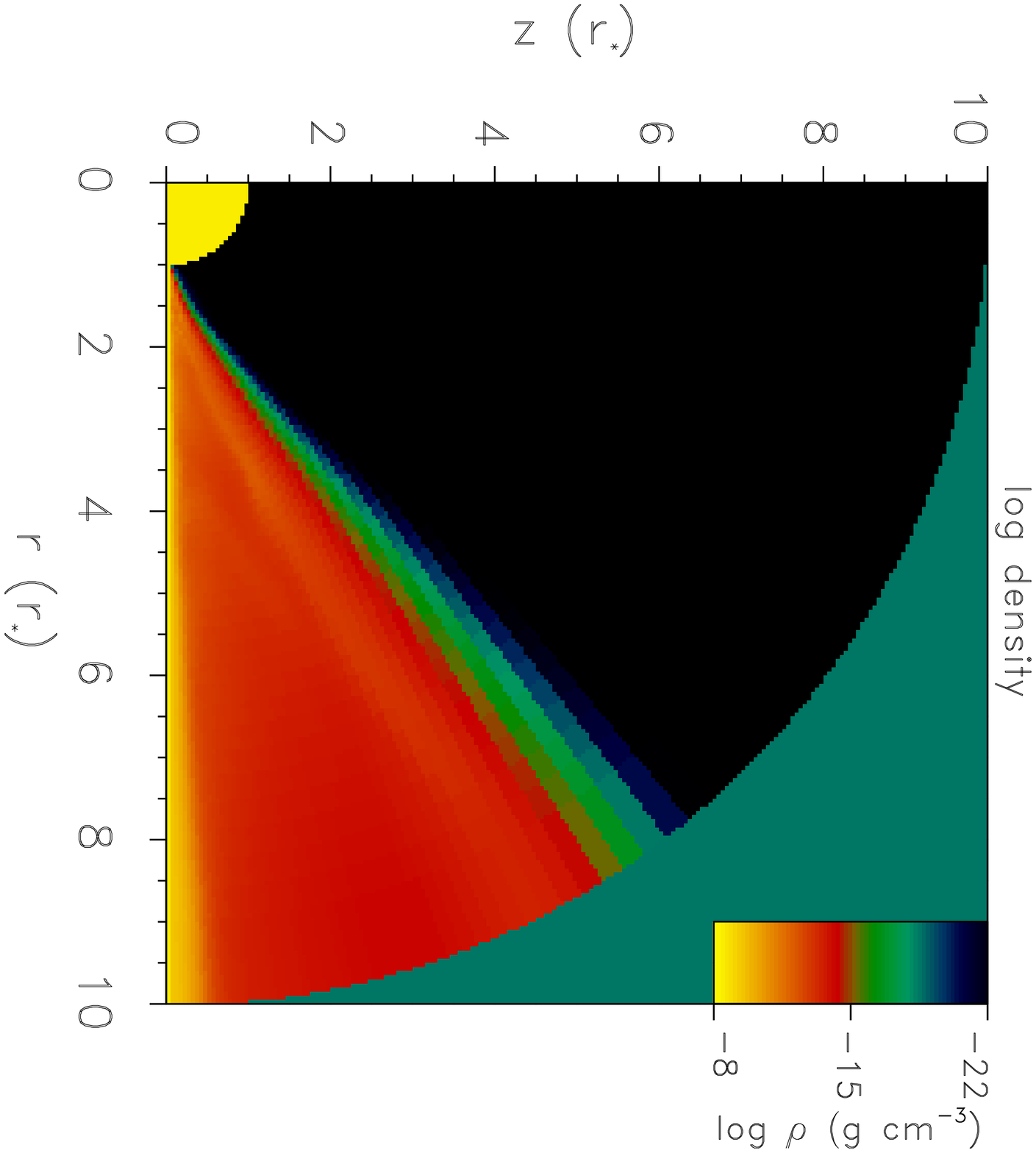}}
\end{picture}
\caption{
The color density maps in two models computed using the method of PSD~II (top
panels), the PSD~II method with the specific angular velocity conservation
(middle panels), 
and the latter with a fixed inclination angle of streamlines 
(bottom panels).
The left column shows the results for the $x=0$ model
while the right column shows results for the $x=1$ model
(see table~2, and section~3.1 for discussion).}
\end{figure}

\eject

\normalsize
\centerline{ Table 1. Summary of results for disc winds }
\begin{center}
\begin{tabular}{c c c c c c } \\ \hline \hline
     &                         &   &                         &                   &             \\
run &  $\MDOT_a$ & x & $\MDOT_w$ & $v_r(10 r_\ast)$ & $\omega$ or $(90^o -i)^{\ast}$ \\
     & (M$_{\odot}$ yr$^{-1}$ ) &   & (M$_{\odot}$ yr$^{-1}$) & $(\rm km~s^{-1})$ & degrees    \\  

\hline

PSD~II     &                         &   &                         &                   &             \\

A    &$  10^{-8}$              & 0 &  $ 5.5\times10^{-14}$     & 900                 &  50  \\
B    &$ \pi \times 10^{-8}$    & 0 &  $ 4.0\times10^{-12}$      & 3500                &  60  \\
C    &$ \pi \times 10^{-8}$    & 1 &  $ 2.1\times10^{-11}$     & 3500                &  32  \\

 & & & & &   \\

I     &                         &   &                         &                   &             \\

a    &$  10^{-8}$              & 0 &  $ 1.3\times10^{-12}$     & $15000$                 &  15  \\

b    &$ \pi \times 10^{-8}$    & 0 &  $ 1.3\times10^{-11}$      & 20000                & 38   \\

c    &$ \pi \times 10^{-8}$    & 1 &  $ 2.4\times10^{-11}$     & 32000                &  12  \\

 & & & & &   \\

II     &                         &   &                         &                   &             \\

a    &$  10^{-8}$              & 0 &  $ 6.3\times10^{-14}$     & 600                 &  90$\dagger$   \\ 

a'    &$  10^{-8}$              & 0 &  $ 6.3\times10^{-13}$     & 1100                 &  60$\dagger$   \\ 

c    &$ \pi \times 10^{-8}$    & 1 &  $ 4.2\times10^{-11}$     & 5000                &  30$\dagger$  \\

 & & & & &   \\

III     &                         &   &                         &                    &             \\

a    &$  10^{-8}$              & 0 &  $ 6.3\times10^{-13}$     & 16000                 &  60$\dagger$   \\ 

c    &$ \pi \times 10^{-8}$    & 1 &  $ 4.2\times10^{-11}$     & 28000                 &  30$\dagger$  \\

\hline
\end{tabular}
\end{center}
$\ast$ 
For all models from PSD~II and our models in case I, 
the last column lists the wind opening angle, $\omega$.
While for models in  cases II and III, 
the last column lists $90^o-i$ (marked with $\dagger$), where $i$ is 
the assumed inclination angle between the poloidal component of the velocity
and the normal to the disk midplane; note that then $\omega \approx (90^o -i)$.

\end{document}